\documentclass[10pt]{emulateapj}

\shorttitle{Astrometric companions to nearby stars}
\shortauthors{Tokovinin, Hartung, Hayward}

\begin{document}

\title{Companions to nearby stars with astrometric
  acceleration. II. \footnotemark[1]}

\footnotetext[1]{Based on  observations obtained at  the Gemini
    Observatory (Programs GS-2011B-Q69 and GS-20012B-Q-71). }

\author{Andrei Tokovinin}
\affil{Cerro Tololo Inter-American Observatory, Casilla 603, La Serena, Chile}
\email{atokovinin@ctio.noao.edu}

\author{Markus Hartung, and Thomas L. Hayward }
\affil{Gemini Observatory, Southern Operations Center,  Casilla 603, La Serena, Chile}
\email{mhartung@gemini.edu, thayward@gemini.edu}

\begin{abstract}
{\it  Hipparcos}  astrometric binaries  were  observed  with the  NICI
adaptive  optics system at  Gemini-S, completing  the work  of Paper~I
(Tokovinin et al. 2012).  Among the 65 F, G, K dwarfs within 67\,pc of
the  Sun studied  here, we  resolve 18  new  sub-arcsecond companions,
re-measure  7  known astrometric  pairs,  and  establish the  physical
nature  of yet another  three wider  companions.  The  107 astrometric
binaries targeted  at Gemini so  far have 38 resolved  companions with
separations under 3$''$.  Modeling shows that bright enough companions
with separations  on the  order of an  arcsecond can perturb  the {\it
  Hipparcos} astrometry  when they are  not accounted for in  the data
reduction.  However, the resulting  bias of parallax and proper motion
is generally  below formal errors  and such companions  cannot produce
fake  acceleration.    This  work  contributes   to  the  multiplicity
statistics of nearby dwarfs  by bridging the gap between spectroscopic
and visual  binaries and  by providing estimates  of periods  and mass
ratios for many astrometric binaries.
\end{abstract}


\section{Introduction}
\label{sec:intro}

This  work completes  the  study of  astrometric  binaries started  in
\citet[][Paper I]{Paper1}.   Many new companions  to solar-mass nearby
dwarf  stars  were  inferred  from their  accelerated  motion,  either
detected  directly  as $\dot{\mu}$  binaries  by  the {\it  Hipparcos}
mission \citep{HIP1,HIP2}, or by comparing short- and long-term proper
motion \citep[$\Delta \mu$ binaries, see][MK05]{MK05}.  Only a fraction of
those objects  have computed  visual and/or spectroscopic  orbits; the
yet unknown periods of most acceleration binaries range from a few to a
thousand   years.   By   resolving  ``dark''   astrometric  companions
directly, we  get much  tighter estimates of  their periods  (from the
projected separations) and masses (from the apparent magnitude). 

There are 343  stars with accelerated proper motion  (PM) among dwarfs
of  spectral types  F and  G  located within  67\,pc of  the Sun  (the
FG-67pc sample).   In Paper~I,  51 of those  stars were  observed with
adaptive optics  (AO), resolving for  the first time  17 sub-arcsecond
companions and  7 wider companions. The ``success  rate'' was slightly
less  than  expected from  the  binary  statistics.   However, it  was
established that  some acceleration  solutions in the  {\it Hipparcos}
catalog  are spurious.   Some  astrometric companions  could be  white
dwarfs  (WDs), too  faint to  be resolved.   These  two considerations
bring the resolution rate in better agreement with the expectations.

Here we report AO observations  of 65 targets. They are selected among
dwarfs  within  67\,pc  with  color  index  $0.5  <  V-I  <1.0$  which
corresponds to  spectral types  from F5V to  K2V.  Seven of  them were
resolved previously and are  re-observed here for confirmation and for
detection of  orbital motion.  Out  of the 5 tentative  resolutions in
Paper~I,  we confirm  three and  refute  two.  To  our knowledge,  the
remaining  58 stars  were observed  with AO  for the  first  time.  We
resolved  18  sub-arcsecond companions  and  three  wider pairs.   The
observations are  presented in Section~2.  In  Section~3 the influence
of  faint, unrecognized  companions on  {\it Hipparcos}  astrometry is
investigated.   Then in  Section~4 we  discuss the  statistics  of the
combined sample of 107 astrometric binaries observed at Gemini.

\section{Observations and results}
\label{sec:AO}

\begin{figure*}[ht]
\epsscale{1.0}
\plotone{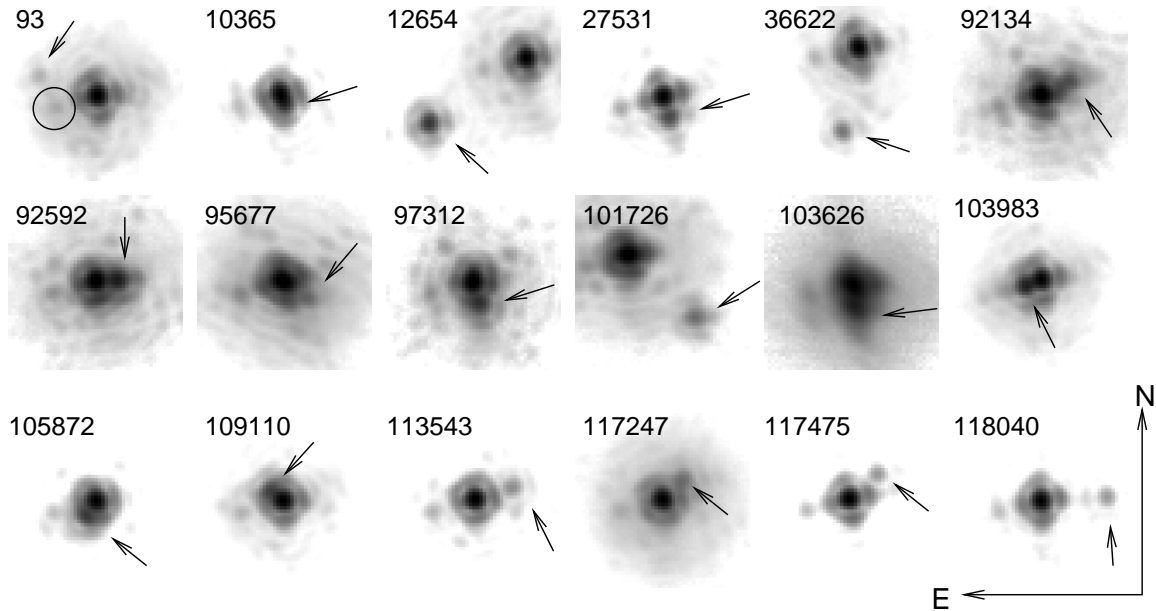}
\caption{Images of  18 newly resolved sub-arcsecond  companions in the
  red channel  (2.272\,$\mu$m), marked  by the HIP  numbers.  Negative
  logarithmic intensity  scale (white is 0.003  of maximum intensity),
  each fragment is 50x50 pixels ($0.9''$).  The ``ghost'' companion to
  the left of each target (circled  in the first image) is a reflex in
  the NICI optics.
\label{fig:images}
}
\end{figure*}

The  Near-Infrared Coronagraphic  Imager  (NICI) on  the Gemini  South
telescope is  an 85-element curvature adaptive  optics (AO) instrument
based on  natural guide stars  \citep{NICI,Chun08}. As in  Paper~I, we
used  NICI  in  normal  (non-coronagraphic)  mode,  with  simultaneous
imaging at two  wavelengths.  The two detectors of  NICI have $1024^2$
pixels of  18\,mas (milliarcseconds) size, covering a  square field of
$18''$.   To avoid  saturation, we  selected narrow-band  filters with
central wavelengths of 2.272\,$\mu$m and 1.587\,$\mu$m for the red and
blue imaging channels.

The  program started  in Paper~I  was continued  by observing  18 more
stars from  the original list  in 2012A, in  the period from  March to
May.  The  observations of  another 47 Hipparcos  astrometric binaries
were taken  in queue mode in  the period from September  2 to December
24, 2012  using 8\,h of  allocated time.  The observing  procedure and
data reduction are  the same as in Paper~I  and in \citet{THH10}.  The
images of  each target at  five dither positions  were median-combined
after removing bad pixels,  subtracting the median sky frame, dividing
by a  flat field, and accounting  for the dither by  shifting back the
images to co-align the source with the first frame.

The total of 29 companions with separations from 0\farcs05 to 9\farcs2
were resolved, 18 of those for the first time.  All resolutions appear
secure (Figure~\ref{fig:images}).  The  reality of detections is checked
by ``blinking''  the red and blue  images and by  comparing with other
stars.  Some companions  are better seen in the  blue images where the
speckle  structure is less  prominent and  point sources  are sharper.
The faint ``ghost'' with $\Delta m  \sim 4.3$ at 0\farcs24 to the left
of each star is produced by the NICI optics.

The limiting magnitude for companion detection was determined from the
intensity fluctuations in annular zones, as in Paper~I.  The detection
depth  depends on  the  AO compensation  quality  which was  variable,
reflecting the seeing variation. The median detection depth in the red
channel is $\Delta  m = 5\fm1$ at 0\farcs27 and $\Delta  m = 7\fm4$ at
0\farcs90.   These   formally  computed  detection   limits  are  only
indicative  because actual detections  depend on  companion's location
and on details of the speckle structure.

Table~\ref{tab:measures} lists  the relative astrometry and
photometry of resolved pairs measured independently on the red and
blue images.  Previously resolved pairs are marked by R in the last
column, the uncertain measures are marked by colons.  For
well-resolved ($\rho > 0\farcs5$) companions the measure is obtained
by fitting the shifted and scaled image of the main companion which
serves as Point Spread Function (PSF).  For closer companions we used
a blind deconvolution.  It starts from the initial estimate of the
binary parameters (separation $\rho$, position angle $\theta$, and
intensity ratio) derived by ``clicking'' on the companion in the
image display.  The halo of the primary companion at $(\rho, \theta +
90^\circ)$ is subtracted from the intensity of the secondary peak.
The first PSF estimate is then obtained by de-convolving the image
from the binary.  At the first iteration, the PSF at distances beyond
$0.75 \rho$ is replaced by its azimuthal average.  This ``synthetic''
PSF is then used for the least-squares fitting of the binary
parameters.  The fitting is done in the Fourier space at spatial
frequencies from $0.2 f_c$ to the cutoff $f_c$, thus neglecting
large-scale intensity variations in the halo and fitting only fine
structure -- the PSF core and speckles around it.  The process is
repeated iteratively (new PSF from the binary, new binary parameters,
etc.)  until convergence, when the rms deviation between the image and
its model does not decrease anymore.  The  reliability is
evaluated qualitatively by the PSF that should have no traces of the
companion. Blind deconvolution works very well in most cases, but it
does not produce reliable results for the faintest or closest
companions near the detection limit.  In such cases the difference
between measures in the red and blue channels informs us of their
quality.

The relative  position of components  in pixels is transformed  to the
position on the sky using the nominal NICI parameters: the pixel scale
of 18.0\,mas and the known offset in position angle.

The  list of  targets observed  in  2012 is  given in  Table~2. It  is
similar to the  Table~2 of Paper~1 and provides  rounded values of the
parallax  $p_{\rm HIP}$,  accelerations $\Delta  \mu$  and $\dot{\mu}$
from \citep{MK05},  radial velocity (RV) variation if  it is variable,
mostly from  the Geneva-Copenhagen  Survey (GCS) of  \citet{N04}. Then
follow the  estimates of the  primary mass $M_1$  and mass ratio  $q =
M_2/M_1$ derived  from known distance, combined  $K_s$ magnitudes from
2MASS \citep{2MASS}, magnitude differences in the red channel measured
here, and the  standard relations of \citet{HM93}. The  values of $q$,
separation  $\rho$, and order-of-magnitude  period estimate  $P^*$ are
given only for  resolved pairs.  The following two  columns of Table~2
list the  detection limits at  15 and 50 pixel  separations (0\farcs27
and 0\farcs90)  in the red  channel (see above).  Notes  on individual
objects are  assembled at the end  of  Table~2.   For some objects,
preliminary  spectroscopic  orbits   were  determined  at  Center  for
Astrophysics  (CfA) in Harvard.\footnote{Latham,  D. W.  2012, private
  communication.}

We  re-observed all  5 binaries  for which  Paper~I  reports tentative
resolutions. Three  (HIP~21778, 22387,  25148) are confirmed  here. Of
the  two  unresolved pairs,  HIP~12425  is  likely  single \citep[no
  acceleration in][]{HIP2},  while HIP~114880 could  have closed in,
considering its short estimated period and variable RV.

The case  of HIP~12654  is perplexing.  A  companion so bright  and so
separated  (0\farcs6) should  have been  resolved by  {\it Hipparcos},
unless  the pair  was much  closer 21  years ago.   If  its semi-major
orbital  axis is  half the  actual separation,  0\farcs3,  the orbital
period should be 60\,yr, so a periastron passage in an eccentric orbit
could  have  happened  during   the  {\it  Hipparcos}  mission.   This
assumption    is    supported     by    $\Delta    \mu    =    (-13.0,
+14.4)$\,mas~yr$^{-1}$,  directed   at  $264^\circ$  (away   from  the
companion) and amounting to 0\farcs4  over 21\,yr.  A few other bright
and well-resolved  companions were  found in Paper~I,  e.g.  HIP~24336
and 21079.  The  effect of binary companions {\it  detectable} but not
actually  {\it detected}  by {\it  Hipparcos} is  studied in  the next
Section.

\section{Effect of faint companions on the Hipparcos astrometry}
\label{sec:comp}

The {\it Hipparcos} satellite obtained the astrometric parameters of
the target stars through one-dimensional scans at different angles
while the spacecraft was spinning.  The star light was modulated using
a grid of slits placed in the focal plane with a slit width of
0\farcs46 and a period of $s = 1\farcs2074$.  The positions of the
stars in scan direction, the {\it abscissae}, were determined by
fitting the first and second Fourier harmonics of the grid period to
the photon counts.   The relative amplitude and phase of the two
  harmonics, $\mu$ and $\nu$, were calibrated for single stars; they
depend slightly on the field position and star color. Significant
deviations from these calibrations were used to detect resolved
binaries with the limiting magnitude difference of $\Delta Hp \la
4^m$.  For the remaining unresolved stars, including all stars in this
program, the number of free parameters was reduced to three by fixing
$\mu$ and $\nu$ to their calibrated values.  Thus, the photon counts
as a function of the scan phase $x$ (in radians) were fitted by a sum
of two sine terms \citep[equation 5.5, p. 54, in][]{Vol3}
\begin{equation} 
I(x) = r_1 + r_2 \cos (x + r_3) + \mu r_2 \cos 2(x + r_3 + \nu)
\label{eq:scan}
\end{equation}
with fixed $\mu$ and $\nu$ and  three free parameters $r_i$.

\begin{figure}[ht]
\epsscale{1.0}
\plotone{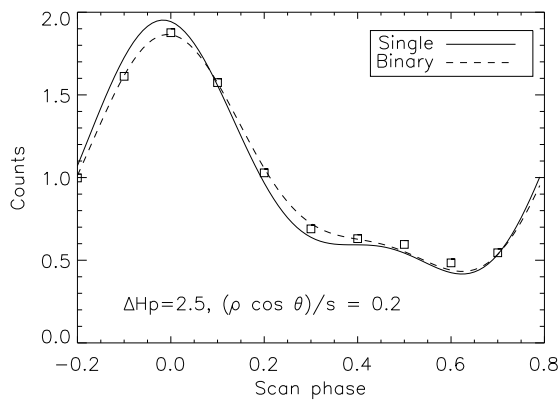}
\caption{Photon counts  vs. scan phase  (fraction of the  grid period)
  for  a single  star (full  line) and  a binary  (dashed  line).  The
  squares  denote a single-star  scan model  fitted to  the scan  of a
  binary.
\label{fig:compmod}
}
\end{figure}

Figure~\ref{fig:compmod} shows a representative  scan of a single star
with  typical  parameters $r_2/r_1=0.7$,  $\mu  =  0.37$,  and $\nu  =
10^\circ$, assuming $r_1 = 1$ in equation~\ref{eq:scan}. The effect of
a binary companion with $\Delta Hp =2.5^m$ and 0\farcs24 separation in
the  scan direction is  shown by  the dashed  line.  Such  a companion
would actually  be recognized  by {\it Hipparcos},  but a star  with a
fainter, un-detected  companion would be treated as  single and fitted
by  the  calibrated  3-parameter   model.   Such  a  fit  (squares  in
Figure~\ref{fig:compmod})   results   in   the  decreased   modulation
amplitude $r_2$ and some shift  of the abscissa (change of $r_3$).  To
investigate the shift of the abscissa caused by undetected companions,
we  fitted single-star  models to  simulated scans  of  binaries, with
uniform weights.  The  shift of the abscissa (i.e.   the star position
in  the scan  direction) $\Delta  v$ can  be represented  by  two sine
terms,
\begin{equation} 
\Delta v \approx s \alpha \; ( 0.103 \sin y + 0.028 \sin 2 y), 
\label{eq:dv}
\end{equation}
where  $\alpha  =  10^{-0.4 \Delta  Hp}$  is  the  flux ratio  of  the
companion in the  {\it Hipparcos} bandpass and $ y =  2 \pi (\rho \cos
\theta)/s$  is the combination  of the  binary separation  $\rho$, its
angle $\theta$  relative to  the scan direction,  and the  grid period
$s$.  The maximum effect of $\Delta v = 0.11 \alpha s$ is reached when
$\rho \cos \theta = 0.2 s  = 0\farcs24$, e.g $\Delta v = 1.3$\,mas for
a companion  with $\Delta Hp =  5$.  When the  projected separation is
$s/2$, the abscissa is not  affected, the companion only decreases the
modulation.   The  coefficients  of  the  model  (\ref{eq:dv})  depend
slightly  on the  weights used  to fit  the scans.   If the  weight is
inversely proportional  to the flux (as  would be the  case of Poisson
noise and  negligible background),  the coefficients become  0.114 and
0.022.

\begin{figure}[ht]
\epsscale{1.0}
\plotone{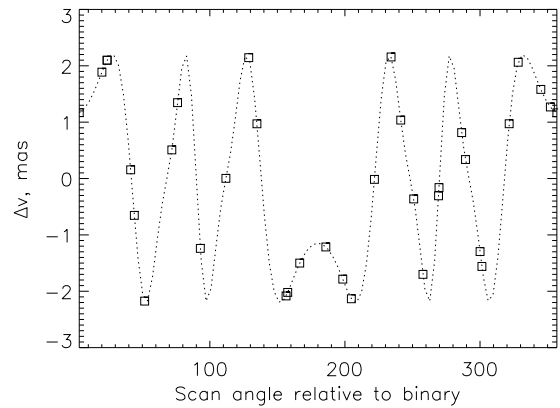}
\caption{Shifts  of great-circle  abscissae $\Delta  v$ caused  by the
  companion of  HIP~21079 (1\farcs622, 23.4$^\circ$)  assuming $\Delta
  Hp = 4.5$, as a function  of the scanning angle.  The dotted line is
  equation  \ref{eq:dv},  the  squares  correspond to  the  36  orbits
  (i.e. scans) used in the astrometric solution for this star.
\label{fig:hip21079}
}
\end{figure}

The five  astrometric parameters of single  stars ($\alpha$, $\delta,$
$\mu_\alpha$, $\mu_\delta$, $\pi$) are derived by fitting the measured
abscissae.  An   undetected  companion  changes   the  abscisssae  and
therefore is  expected to affect the astrometry.   Can these abscissae
shifts explain the $\Delta \mu$  or $\dot{\mu}$ for some wide binaries
resolved by  NICI but treated as  single stars by  {\it Hipparcos}? We
study three case examples to answer this question.

{\it Case 1}: A companion to HIP~21079 was found in Paper~I at $\rho =
1\farcs622$, position angle $23.4^\circ$,  and $\Delta K = 1.62$.  The
Hipparcos  Intermediate  Astrometry Data  (HIAD)  for  this star  were
retrieved        from         the        ESA        archive.\footnote{
  \url{http://www.rssd.esa.int/index.php?project=HIPPARCOS\-\&page=Intermediate\_astrometric\_data}}
For the  36 {\it Hipparcos}  orbits processed by the  FAST consortium,
the  abscissae were  modified  by  adding the  $\Delta  v$ terms  from
equation~\ref{eq:dv}.  We  use the standard relations  for dwarf stars
to estimate  $\Delta V  = 4.5$  and set $\Delta  Hp \approx  \Delta V$
(relative  photometry of  this  pair  at SOAR  in  February 2013  gave
$\Delta V = 4.6 \pm 0.1$).   The scanning direction for each orbit was
calculated from the  derivatives $\partial v / \partial  \alpha *$ and
$\partial    v   /    \partial    \delta$   given    in   the    HIAD.
Figure~\ref{fig:hip21079}  shows  the resulting  $\Delta  v$, with  an
amplitude of  $\pm 2.2$\,mas.  The rms abscissa  residuals caused only
by the effect  of the companion, i.e.  ignoring  any other measurement
errors, are 1.4\,mas.  When we  fit five astrometric parameters to the
modified abscissae by unweighted least-squares, the residuals decrease
only  slightly,  to 1.2\,mas  (the  actual  rms  abscissa residual  is
5.1\,mas).  The  parallax does  not change, the  PM changes  by $(0.3,
-0.4)$\,mas~yr$^{-1}$.  The  effect of the companion  is therefore too
small to explain $\Delta \mu = 5$\,mas~yr$^{-1}$ for this star.  As we
see in  Figure~\ref{fig:hip21079}, the  abscissa shifts caused  by the
companion  are quasi-random, they  oscillate in  function of  the scan
angle.  These  shifts do not  correlate significantly with any  of the
five astrometric parameters and,  therefore, cannot be ``absorbed'' by
modifying the single-star  astrometric solution.  The estimated period
of  this  binary,  570\,yr  (Paper~I),  makes  it  unlikely  that  the
companion  moved substantially  in  21\,yr since  the {\it  Hipparcos}
mission.

{\it  Case  2:} For  HIP~12654  ($\rho  =  0\farcs60$, position  angle
$125.0^\circ$, $\Delta K  = 1.03$) we did the  same analysis as above.
If its companion is a normal  main sequence dwarf we expect $\Delta Hp
\approx  1.8$.   Here, the  rms  shifts  of  abscissae caused  by  the
companion are 12.4\,mas before  adjusting the astrometric solution and
11.0\,mas after, while  the actual rms residual of  the FAST abscissae
is 4.1\,mas.   The effect of the companion  is too strong  compared to the
actual residuals. However,  it is likely that at the  time of the {\it
  Hipparcos}  mission (1991.25)  the  companion of  HIP~12654 was  too
close to be resolved, while causing  the real PM effect of $\Delta \mu
=19$\,mas~yr$^{-1}$.

The alternative  reductions by the  NDAC consortium give  very similar
results  for both  HIP~21079 and  HIP~12654,  and the  new catalog  of
\citet{HIP2}   does  not  differ   significantly  from   the  original
reductions.

{\it Case 3:} HIP~103260 is the  known binary I~18 with a companion of
$\Delta V = 3.2$ according to the WDS \citep{WDS}.  This companion was measured in
Paper~I  at 3\farcs975 and  $351.3^\circ$, but  was not  recognized by
{\it Hipparcos}, which gives  an acceleration solution with $\dot{\mu}
=  7$\,mas~yr$^{-2}$.   The  26  orbits  covering this  star  are  not
distributed in time uniformly and tend towards the end of the mission,
creating correlation between PM  and acceleration in the least-squares
fitting of the  abscissae.  The resulting ill-conditioned acceleration
solution amplifies the noise,  including the abscissa shifts caused by
the  companion  (5.0\,mas rms).  This  effect  has  been discussed  in
Paper~I;  spurious acceleration of  11\,mas~yr$^{-2}$ was  obtained by
fitting a 7-parameter solution  to the companion-induced $\Delta v$ of
this binary.   The unrecognized companion also biases  the parallax by
+2.8\,mas  in  the  5-parameter  solution  and  by  +1.6\,mas  in  the
7-parameter  acceleration  solution.   Analogous  distortion  of  {\it
  Hipparcos}  parallaxes by  {\it close}  companions was  evidenced by
\citet{ST98}. In  this case the  shifts $\Delta v(\theta)$  resemble a
sine wave and correlate stronger with the astrometric parameters.

In summary,  faint undetected  binary companions typically  modify the
measured   abscissae  by  a   few  mas.    As  these   ``errors''  are
quasi-random,  their effect  on the  standard  5-parameter astrometric
solution is  usually below  its formal errors,  except for  cases like
HIP~103260 (Case 3) where its parallax or PM can be slightly biased or
when  an  ill-conditioned  acceleration  solution  amplifies    all
(measurement and companion-induced) errors.

\section{Statistics  and discussion}
\label{sec:stat}

\begin{figure}[ht]
\epsscale{1.0}
\plotone{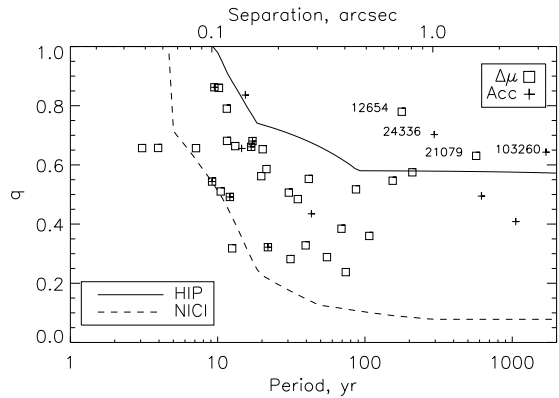}
\caption{Detection limits  of NICI and Hipparcos in  the (period, mass
  ratio) parameter  space and resolved astrometric  binaries from this
  work and  Paper~I. Four  pairs with long  periods and large  $q$ are
  labeled by their HIP numbers.
\label{fig:det}
}
\end{figure}

Paper~I evoked published speckle interferometry to study the
statistics of resolved astrometric binaries in a sample of 99
objects. Here we use only the NICI data (which go deeper than speckle)
and discuss the merged sample of 107 objects observed at Gemini. 
  The estimated periods $P^*$ and mass ratios $q$ of these pairs are
plotted in Figure~\ref{fig:det}.  The full and dashed lines depict the
typical detection limits of {\it Hipparcos} and NICI, respectively, at
a distance of 50\,pc. The two pairs with the shortest periods around
3\,yr (HIP~10365 and 109110) are formally beyond the limit. There are
38 companions with $\rho < 3''$ ($P^* \la 2000$\,yr) -- 20 from
Paper~I and 18 from this work.\footnote{Although the resolution of
  HIP~114880 is not confirmed here, we consider it real because of the
  variable RV.}

The detection rate of $38/107 = 35.5 \pm 6$\% is significantly below
the 55\% rate predicted for NICI by the simulations of Paper~I.  As in
Paper~I, we correct the observed rate to 45\% by accounting for 10\%
of known binaries that were excluded from the NICI program and for
10\% of WD companions, but the result is still too low.  Remember that
the simulations are sensitive to the assumed binary statistics and
contain some simplifications.  The observed detection rate is
uncertain because of hierarchical multiples (some resolutions are
actually tertiary companions where the inner closer sub-systems
produce the acceleration).  For these two resasons we are not
  concerned about  the remaining disagreement between 
  observed and simulated detection rates.  If we ignore $\dot{\mu}$
binaries and consider only 83 $\Delta \mu$ binaries where 30
companions with $P^* \le 100$\,yr are resolved, the observed rate is
$36.1 \pm 6.5$\%.

Note the striking absence of  resolved companions with $P^* > 100$\,yr
and  $q <  0.4$  in the  lower-right  corner of  Figure~\ref{fig:det},
despite their easy detectability with NICI (one such pair was resolved
with  speckle, see  Figure~3b of  Paper~I).   There is  no doubt  that
low-mass  companions  with  $P  >100$\,yr  are  abundant  \citep{R10}.
However, at such long periods  the $\sim$100\,yr time baseline of {\it
  Tycho-2} samples only a fraction  of the orbit and the {\it Tycho-2}
PM no longer  represents the center-of-mass motion, as  assumed in our
simulations  described in  Paper~I.  Therefore  the $\Delta  \mu$ (the
difference between  {\it Hipparcos} and {\it Tycho-2}  PMs) is reduced
(both catalogs measure nearly the same) and the chances of discovering
$\Delta \mu$ are reduced.
If  most  $\Delta  \mu$  binaries  indeed have  periods  shorter  than
100\,yr,  their  number  decreases   by  $\sim$18\%  compared  to  our
simulations (see the cumulative curve in Figure~4a of Paper~I).

The  NICI  companions  with  $P^*  >  100$\,yr  and  relatively  large
estimated   $q$   (labeled  points   in   the   upper-right  part   of
Figure~\ref{fig:det}) were  not resolved  by {\it Hipparcos}  and were
treated as single stars. They  are too wide to cause real $\dot{\mu}$.
Our modeling shows that in most cases such companions do not produce a
substantial effect on  astrometry, unless the companion-induced errors
are amplified in  ill-conditioned acceleration solutions (HIP~103260).
Either those $\dot{\mu}$ are spurious,  or we deal with triple systems
where the  acceleration is  produced by inner,  unresolved sub-systems
(HIP~16853 and 109443 have variable RV).  On the other hand, HIP~12654
and 21079  have standard 5-parameter solutions, their  $\Delta \mu$ is
most likely real and caused by their wide companions.

Some secondary  companions could be close binaries,  like in HIP~11072
\citep{11072}.  In  this case  they are redder  than single  dwarfs of
equal luminosity,  making their  direct resolution by  {\it Hipparcos}
and  NICI  more difficult,  while  the  large  combined mass  produces
detectable  astrometric  accelerations.   Presence  of such  dark  and
massive companions in the  binary population improves the agreement of
simulations with reality (Paper I).

Low-mass  companions also  produce color-dependent  shifts  of stellar
positions.  The companion of HIP~21079 with $\Delta K = 1.9$ and $\rho
= 1\farcs6$ displaces the image centroid in the $K$ band by 0\farcs25 in
the direction  of $23^\circ$. The difference  of equatorial coordinates
of this star between 2MASS and {\it Hipparcos} (both for the epoch and
equinox  J2000.0) implies  a displacement  of 0\farcs14  at $30^\circ$
angle  and  matches   the  companion's  angle.   However,  considering
systematic   errors,    uncertainties   of   the    2MASS   astrometry
($\sim$60\,mas)  and the  PM errors  multiplied by  the  difference of
epochs, this effect could not  be accepted as a companion detection if
the latter were not directly  resolved with NICI.  In conclusion, very
precise simultaneous astrometry at  different wavelengths can reveal a
companion's presence and can be used for their detection.

During  few   nights  allocated  to  this  project   we  surveyed  107
acceleration binaries out  of few thousand reported in  MK05, or about
1/3 of  343 such objects in  the FG-67pc sample.   Therefore, no major
changes  of our  statistical  results  are expected  and  as the  NICI
instrument is no longer available, we consider this work as completed.

The  observations reported here  and in  Paper~I explore  nearby dwarf
binaries  in the  regime  difficult  to study  by  other methods:  the
periods are  mostly too long for  RV coverage, the  companions are too
faint to be resolved  at visible wavelengths. Astrometric detection of
such binaries by {\it Hipparcos} is therefore an excellent opportunity
to  bridge  the  gap   between  spectroscopy  and  imaging.   Improved
characterization  of  astrometric   binaries  and  understanding  some
caveats of  {\it Hipparcos}  astrometry are the  main results  of this
study.  In  a broader perspective, it contributes  to the multiplicity
statistics of nearby dwarfs, many  of which are targeted by exo-planet
programs.

\acknowledgments We thank Fredrik  Rantakyro and Peter Pessev (Gemini)
for  careful  observations of  our  program  stars,  David Latham  for
sharing  unpublished data  from  the CfA  survey,  Valeri Makarov  for
discussion of  companion's impact  on {\it Hipparcos}  astrometry, and
Wolfgang  Brandner  for  general  remarks  on  binary  statistics  and
methodics.  This work  used the SIMBAD service operated  by Centre des
Donn\'ees  Stellaires (Strasbourg,  France),  bibliographic references
from  the  Astrophysics  Data  System  maintained  by  SAO/NASA,  data
products of the  Two Micron All-Sky Survey (2MASS)  and the Washington
Double Star Catalog maintained at USNO.
Gemini telescopes  are  operated by the  Association of Universities
for Research  in Astronomy, Inc.,  under a cooperative  agreement with
the  NSF on  behalf of  the Gemini  partnership: the  National Science
Foundation  (United  States), the  Science  and Technology  Facilities
Council  (United  Kingdom), the  National  Research Council  (Canada),
CONICYT   (Chile),  the   Australian  Research   Council  (Australia),
Minist\'{e}rio da Ci\^{e}ncia e  Tecnologia (Brazil) and Ministerio de
Ciencia, Tecnolog\'{i}a e Innovaci\'{o}n Productiva (Argentina).

{\it Facilities:} \facility{Gemini:South (NICI)}


\begin{deluxetable*}{l c rcc rcc l}            
\tabletypesize{\scriptsize}                                                                                                               
\tablecaption{Measures of  resolved companions
\label{tab:measures}   }                                                                                                                       
\tablewidth{0pt}                                                                                                                          
\tablehead{  HIP   & Date & \multicolumn{3}{c}{Red 2.272\,$\mu$m} &  \multicolumn{3}{c}{Blue 1.587\,$\mu$m} &  Rem \\ 
                   &  &   P.A.         &  Sep.   & $\Delta m$ &     P.A.         &  Sep.   & $\Delta m$ &  \\
                   &  & (deg)          & (arcsec)    & (mag)      &   (deg)          & (arcsec)    & (mag)      & }
\startdata                                                                                                       
93    & 2012.8317 &       71.8 &    0.317 &     4.34 &       72.4 &    0.323 &     4.34 &  \\
1103  & 2012.7361 &      302.0 &    6.193 &     4.47 &      302.5 &    6.204 &     4.63 &  \\
6273  & 2012.6707 &      113.6 &    0.254 &     2.39 &      113.5 &    0.254 &     2.71 & R \\
6712  & 2012.7362 &       28.8 &    0.100 &     0.84 &       29.9 &    0.101 &     0.97 & R \\
10365 & 2012.6704 &      198.4 &    0.056 &     0.83 &      198.2 &    0.054 &     0.47 & : \\
11072 & 2012.6706 &      339.3 &    0.345 &     1.78 &      339.9 &    0.344 &     1.53 & R \\
12654 & 2012.7362 &      123.8 &    0.610 &     1.09 &      123.8 &    0.610 &     1.24 &  \\
21778 & 2012.6706 &      189.4 &    0.148 &     2.98 &      188.5 &    0.143 &     3.42 & R \\
22387 & 2012.6706 &       58.8 &    0.140 &     4.19 &       62.6 &    0.143 &     3.38 & R: \\
25148 & 2012.6707 &      209.4 &    0.065 &     1.21 &      213.0 &    0.059 &     0.80 & R: \\
27531 & 2012.7363 &      196.2 &    0.118 &     1.54 &      196.3 &    0.116 &     1.83 &  \\
36622 & 2012.9745 &      169.2 &    0.445 &     2.44 &      169.1 &    0.446 &     2.71 &  \\
88595 & 2012.2379 &      293.0 &    6.609 &     5.52 &      293.1 &    6.624 &     7.15 &  \\
92134 & 2012.2380 &      294.0 &    0.166 &     2.20 &      295.3 &    0.165 &     2.29 & AB  \\
92134 & 2012.2380 &      292.5 &    9.164 &     3.98 &      292.6 &    9.155 &     4.20 & AC \\
92592 & 2012.2380 &      271.4 &    0.121 &     1.04 &      271.6 &    0.124 &     1.28 &  \\
95677 & 2012.2380 &      232.8 &    0.161 &     3.65 &      234.2 &    0.156 &     3.39 & : \\
97312 & 2012.4132 &      194.1 &    0.118 &     1.61 &      190.7 &    0.112 &     1.61 &  \\
101726 & 2012.2381 &      226.9 &    0.483 &     2.95 &      226.8 &    0.482 &     3.19 &  \\
103626 & 2012.3449 &      188.1 &    0.123 &     0.94 &      188.3 &    0.121 &     1.07 &  \\
103983 & 2012.8313 &      110.0 &    0.087 &     0.60 &      111.6 &    0.087 &     0.71 &  \\
105872 & 2012.8996 &      149.2 &    0.097 &     1.70 &      138.1 &    0.087 &     1.61 &  \\
107731 & 2012.3450 &      307.3 &    5.587 &     3.59 &      307.3 &    5.561 &     4.21 &  \\
109110 & 2012.8315 &       48.5 &    0.082 &     1.74 &       46.9 &    0.080 &     1.73 & : \\
113543 & 2012.8316 &      287.5 &    0.214 &     2.92 &      287.2 &    0.208 &     3.05 &  \\
117247 & 2012.8317 &      312.4 &    0.151 &     2.94 &      309.8 &    0.150 &     2.51 &  \\
117258 & 2012.8316 &       20.0 &    0.227 &     1.96 &       19.7 &    0.226 &     2.24 & R \\
117475 & 2012.8315 &      310.5 &    0.196 &     3.27 &      309.5 &    0.196 &     3.72 &  \\
118040 & 2012.8315 &      273.3 &    0.347 &     3.17 &      273.3 &    0.347 &     3.41 &  
\enddata                                                                                                                       
\end{deluxetable*}

\begin{deluxetable*}{l ccc cc ccc cc   }            
\tabletypesize{\scriptsize}                                                                                                               
\tablecaption{Summary data on observed astrometric binaries
\label{tab:all}   }                                                                                                                       
\tablewidth{\textwidth}                                                                                                                          
\tablehead{  HIP   & $p_{HIP}$ & $\Delta \mu$ & $\dot{\mu}$ & $\Delta$RV &$M_1$ & $q$ & $\rho$ & $P^*$ & $\Delta m_{15}$ & $\Delta m_{50}$  \\
                   & (mas) & (mas~yr$^{-1}$) & (mas~yr$^{-2}$) & (km~s$^{-1}$) & ($M_\odot$) &    & (arcsec) & (yr)   & (mag) & (mag)       }
\startdata    
    93 &   16 &    6 &    0 &    1.2 &   1.08 &   0.24 &  0.317 &   74.1 &   5.26 &   7.15   \\
   290 &   15 &   13 &    0 &    1.3 &   1.19 &        &        &        &   4.97 &   7.32   \\
   305 &   21 &   12 &   10 &    0.0 &   1.10 &        &        &        &   5.39 &   7.80   \\
   359 &   17 &   15 &    0 &    -   &   0.85 &        &        &        &   4.97 &   6.18   \\
  1103 &   15 &    8 &   11 &    -   &   1.20 &   0.25 &  6.193 & 6654.3 &   5.57 &   7.69   \\
  1274 &   15 &   14 &    7 &    -   &   1.01 &        &        &        &   5.53 &   7.40   \\
  1573 &   22 &   13 &    0 &    1.3 &   1.12 &        &        &        &   5.36 &   7.75   \\
  1976 &   21 &   13 &   20 &    5.1 &   1.07 &        &        &        &   5.36 &   7.80   \\
  3578 &   25 &    7 &   15 &    2.4 &   0.94 &        &        &        &   5.29 &   7.81   \\
  4668 &   15 &   10 &    0 &    -   &   0.93 &        &        &        &   5.30 &   7.12   \\
  4981 &   17 &    6 &    0 &    -   &   0.84 &        &        &        &   5.52 &   7.46   \\
  6273 &   30 &   19 &    0 &    2.2 &   0.92 &   0.56 &  0.254 &   19.8 &   5.04 &   7.86   \\
  6712 &   18 &   16 &   25 &    0.8 &   0.94 &   0.82 &  0.100 &    9.3 &   5.28 &   7.70   \\
  7961 &   20 &   11 &    0 &    1.7 &   1.17 &        &        &        &   5.34 &   7.91   \\
  8674 &   19 &   14 &    0 &    0.0 &   0.87 &        &        &        &   5.64 &   7.85   \\
 10365 &   19 &    7 &    0 &    -   &   0.94 &   0.66 &  0.047 &    3.1 &   5.38 &   7.67   \\
 11072 &   45 &    0 &   19 &    0.0 &   1.22 &   0.66 &  0.343 &   14.5 &   1.95 &   5.98   \\
 12425 &   15 &    0 &   17 &    0.0 &   0.99 &        &        &        &   5.55 &   7.93   \\
 12654 &   16 &   19 &    0 &    0.0 &   0.84 &   0.78 &  0.614 &  177.7 &   3.16 &   4.32   \\
 17184 &   20 &   10 &   28 &    0.0 &   0.93 &        &        &        &   4.94 &   7.41   \\
 19248 &   28 &   12 &    5 &    SB  &   0.92 &        &        &        &   5.48 &   7.71   \\
 21778 &   23 &   15 &   11 &    1.6 &   0.98 &   0.49 &  0.141 &   12.1 &   5.65 &   7.79   \\
 22387 &   18 &    9 &    8 &    2.3 &   1.06 &   0.19 &  0.142 &   18.3 &   5.51 &   7.90   \\
 23641 &   24 &    9 &   33 &    4.1 &   0.79 &        &        &        &   5.42 &   7.49   \\
 25148 &   15 &    5 &    0 &    3.7 &   0.97 &   0.75 &  0.065 &    6.8 &   5.23 &   7.38   \\
 27531 &   21 &   14 &    0 &    1.0 &   0.72 &   0.68 &  0.118 &   11.6 &   5.16 &   7.64   \\
 30509 &   17 &    8 &    0 &    2.2 &   1.09 &        &        &        &   5.41 &   7.47   \\
 34212 &   17 &    7 &   23 &    4.3 &   1.13 &        &        &        &   5.13 &   7.12   \\
 34961 &   18 &    9 &   12 &    0.0 &   0.88 &        &        &        &   5.38 &   7.17   \\
 36622 &   20 &    8 &    0 &    0.0 &   0.86 &   0.52 &  0.445 &   86.9 &   2.74 &   7.29   \\
 38134 &   19 &   11 &    6 &    5.6 &   1.05 &        &        &        &   5.38 &   7.66   \\
 88595 &   17 &    0 &    9 &    -   &   0.97 &        &        &        &   5.54 &   7.47   \\
 91215 &   15 &    0 &    8 &    0.0 &   1.19 &        &        &        &   5.39 &   7.64   \\
 92134 &   17 &   14 &    0 &    0.0 &   0.96 &   0.59 &  0.158 &   21.4 &   5.38 &   7.49   \\
 92592 &   19 &    7 &    0 &    -   &   0.87 &   0.79 &  0.115 &   11.5 &   5.04 &   7.49   \\
 94370 &   18 &    5 &    0 &    0.8 &   1.00 &        &        &        &   5.27 &   7.55   \\
 94668 &   16 &   12 &   34 &    2.0 &   1.01 &        &        &        &   5.46 &   7.50   \\
 95677 &   17 &   10 &    6 &    0.0 &   1.18 &   0.54 &  0.096 &    9.2 &   5.38 &   7.55   \\
 95696 &   15 &    7 &   19 &    1.6 &   1.14 &        &        &        &   5.45 &   7.73   \\
 96754 &   17 &    0 &    8 &    0.0 &   1.20 &        &        &        &   5.08 &   7.67   \\
 97312 &   15 &   18 &   18 &    0.0 &   1.08 &   0.68 &  0.123 &   17.2 &   5.31 &   7.53   \\
 98108 &   16 &    0 &   11 &    2.3 &   1.13 &        &        &        &   5.54 &   7.63   \\
 99708 &   15 &    0 &   12 &    -   &   1.01 &        &        &        &   5.47 &   7.16   \\
100934 &   18 &   18 &    8 &    0.8 &   0.96 &        &        &        &   4.61 &   5.22   \\
101726 &   26 &    8 &    0 &    0.0 &   0.85 &   0.38 &  0.475 &   69.3 &   2.82 &   8.20   \\
102130 &   18 &    7 &    0 &    4.8 &   1.04 &        &        &        &   5.42 &   7.48   \\
103626 &   16 &    0 &   15 &    -   &   0.83 &   0.84 &  0.116 &   15.4 &   4.45 &   6.87   \\
103983 &   15 &    9 &    0 &    0.2 &   0.96 &   0.86 &  0.087 &   10.2 &   4.95 &   7.19   \\
105872 &   15 &   10 &    0 &    -   &   0.91 &   0.66 &  0.097 &   13.2 &   5.49 &   7.38   \\
105879 &   15 &    9 &    7 &    2.3 &   1.33 &        &        &        &   4.32 &   6.92   \\
106560 &   24 &    6 &    5 &    1.7 &   0.91 &        &        &        &   5.09 &   6.41   \\
107239 &   16 &    8 &    0 &    -   &   0.92 &        &        &        &   4.94 &   7.17   \\
107731 &   15 &   10 &   12 &    -   &   1.03 &   0.34 &  5.587 & 5831.4 &   4.40 &   6.00   \\
108589 &   17 &    7 &    0 &    8.5 &   1.00 &        &        &        &   5.59 &   7.86   \\
109110 &   28 &   17 &    0 &    SB  &   0.93 &   0.66 &  0.082 &    3.9 &   5.21 &   7.65   \\
109122 &   16 &   15 &   22 &    4.0 &   1.25 &        &        &        &   5.46 &   7.79   \\
110340 &   19 &    7 &    0 &    0.0 &   0.90 &        &        &        &   5.58 &   7.40   \\
112396 &   15 &    7 &    0 &    -   &   0.86 &        &        &        &   5.80 &   7.41   \\
113543 &   17 &   11 &    0 &    SB  &   1.13 &   0.51 &  0.205 &   30.4 &   5.26 &   7.71  \\
114040 &   16 &   14 &   11 &    5.7 &   1.02 &        &        &        &   4.92 &   6.60   \\
114880 &   16 &    8 &    0 &    2.5 &   1.08 &        &        &        &   4.35 &   6.43   \\
117247 &   27 &    8 &    0 &    -   &   0.80 &   0.32 &  0.151 &   12.5 &   5.55 &   6.60   \\
117258 &   25 &   11 &    0 &    1.4 &   1.02 &   0.62 &  0.227 &   20.1 &   4.49 &   7.30   \\
117475 &   15 &    5 &    0 &    -   &   0.99 &   0.33 &  0.197 &   39.6 &   5.65 &   7.75   \\
118040 &   19 &   13 &    0 &    -   &   0.81 &   0.31 &  0.347 &   75.4 &   5.19 &   7.62   \\
\enddata 
\tablecomments{\\  HIP 93:  {\it  first  resolution}.   Spectroscopic
  binary with a long period  acording to GCS and CfA. \\
HIP 305: {\it unresolved}. RV=const, $\Delta \mu=12$\,mas~yr$^{-1}$, possibly single. \\
HIP  1103:  {\it likely triple system}. We confirm the companion at
6\farcs17, 303$^\circ$ from 2MASS as physical, i.e. common proper
motion (CPM). The 2MASS gives the companion's $J-K$ color too blue for
a dwarf; the erroneous photometry is possibly caused by the proximity
of the bright component.  This pair is too wide to cause the
acceleration, so the system is likely triple. The star was on the GCS
survey but has no RV data.  It is a ROSAT X-ray source.  \\
}
\end{deluxetable*}  
\setcounter{table}{1}
\begin{deluxetable*}{l ccc cc ccc  l }        
\tabletypesize{\scriptsize}   
\tablecaption{ (Continued)}
\tablewidth{\textwidth}    
\tablehead{}
\startdata
\enddata
\tablecomments{\\
HIP 3578: {\it unresolved} with NICI and speckle despite acceleration
and variable RV. SIMBAD gives the wrong spectral  type F0IV; it
  is a 1.0 solar-mass dwarf with $V-K=1.63$.  \\
HIP 4981: {\it unresolved}, possibly single: small $\Delta \mu$, no RV
data. \\
HIP 6273: retrograde motion by $4^\circ$ since its first resolution at
Gemini on  2011.84.  Also  measured at SOAR  with speckle  on 2012.93.
\citet{GM06}  propose  two  orbits  with  periods  around  9\,yr,  the
semi-major  axis should  then  be  0\farcs15, so  the  system is  near
apastron.  It will close down and move faster in the coming years. \\
HIP 6712: direct motion by 20$^\circ$ since first resolution at Gemini
on 2011.84 at 9$^\circ$, 0\farcs10, at constant separation.  Estimated
period   about  10\,yr.   Also   measured  at   SOAR  on   2012.92  at
39.0$^\circ$, 0\farcs0998.   The small RV amplitude  suggests an orbit in
the plane of the sky. \\
HIP 7961: {\it unresolved astrometric  and spectroscopic binary.} The visual
companion BUP 24 at 85\arcsec is likely  optical because it is not recovered in
2MASS and has only one measure in the WDS. \\
HIP  8674: unresolved.  Possibly  single with  a  small $\Delta
\mu=12$\,mas~yr$^{-1}$ and constant RV. \\
HIP 10365: {\it first resolution.} A short 3-yr period is expected but
no   RV  data   exist.    Strangely,  {\it   Hipparcos}  measured   no
acceleration, only $\Delta \mu=7$\,mas~yr$^{-1}$. The pair is below the
formal detection limit, but it is resolved securely. \\
HIP  11072:  astrometric  binary  with large  acceleration  where  the
massive   companion  B   is  in   fact  a   close  pair   of  M-dwarfs
\citep{11072}. The image of the primary is strongly saturated. For this
reason the measurements reported in Table~1 are obtained by PSF-fitting at
radii from  5 to  10 pixels to  avoid the center.   The relative
photometry is  uncertain. \\
HIP 12425:  tentative resolution on 2011.84  at 78$^\circ$, 0\farcs34,
$\Delta  K=3.8$   is  not   confirmed  here,  with   new  good-quality
images. The star  has a constant RV, so it could  be single.  In fact,
the    17\,mas~yr$^{-2}$   acceleration    is    not   confirmed    by
\citet{HIP2}. \\
HIP 12654: the 0\farcs6 pair with $\Delta K=1.1$ is obvious and should
have been resolved both by visual observers and by {\it Hipparcos}.
Yet it is not listed as binary in the WDS \citep{WDS} and no
indications of previous resolution are found in the literature.  The
declination of $-79^\circ$ may have something to do with the missed
companion as the southern sky is less well surveyed for binaries. \\
HIP 19248: {\it triple system} consisting  of a 2.5-d spectroscopic pair and
a  0\farcs1   tertiary  companion  discovered   in  \citep{Tok06}  and
previously revealed by the  {\it Hipparcos} acceleration and RV trend.
It is not resolved here, apparently it closed in. \\
HIP  21778: direct  motion  by 16$^\circ$  since  first resolution  at
Gemini  on 2011.80 (176$^\circ$,  0\farcs165), closing  down. Estimated
period 15\,yr. \\
HIP 22387: retrograde motion  by 5$^\circ$ since resolution on 2011.78
at  65$^\circ$, 0\farcs16.   Period  about 20\,yr.   The measurements  are
uncertain because of the large  $\Delta m$.  The ``blue'' color of the
companion ($\Delta H < \Delta K$) is caused by measurement errors. \\
HIP 25148: the tentative resolution of this close pair at Gemini on
2011.70 at 195$^\circ$, 0\farcs06 is confirmed here with 25$^\circ$ of
direct motion.  Again, the $\Delta H < \Delta K$ likely results 
  from the measurement errors. \\
HIP 27531: {\it first resolution,} $P^*=12$\,yr.  \\
HIP  34212:  the  image seems elongated  at  45$^\circ$, {\it partially
resolved?}  A large  RV amplitude and large acceleration  hint at a short
period. \\
HIP 36622: {\it first resolution,}  $P^*=85$\,yr.  \\
HIP 38134: this metal-poor astrometric and spectroscopic binary is not
resolved here. \\
HIP  88595: the  faint and  red  companion at  6\farcs6, 293\degr  ~is
likely optical. This  is a very crowded field in  the direction of the
Galactic center. We see another  companion at $10.8''$, but only in the
red channel. The acceleration might be spurious. \\
HIP 91195: likely spurious acceleration, constant RV. \\
HIP  92134:  {\it  triple  system.}  The $\Delta  \mu$  binary  AB  is
resolved.  The wider  companion C  (9\farcs16,  292.5$^\circ$, $\Delta
K=3.98$) is confirmed as CPM  by 2MASS (8\farcs9, 292.8\degr), it also
has a $J-K$ color corresponding to its estimated mass.  The measurement of
AC   by  PSF   fitting  is   not  strongly   affected  by   the  faint
B-companion. \\
HIP  92592:  {\it triple  system.}  The  inner  astrometric binary  is
resolved   here,  the   CPM   companion  at   146$''$   is  found   by
\citet{Lepine2012}. \\
HIP   94370:   unresolved.    $\Delta   \mu$  and   long-period
spectroscopic binary according to CfA. \\
HIP 94668: variable RV and large acceleration, close binary? \\
HIP 95677: {\it first resolution,} $P^* = 9$\,yr. \\
HIP 96754:  this double-lined  binary with  $q=0.822$ according  to GCS is
apparently too close for resolution with NICI. \\
HIP 97312: {\it first resolution,} $P^*=17$\,yr, but constant RV according to GCS. \\
HIP 98108: acceleration and spectroscopic binary, unresolved here. \\
HIP 99708: no RV data, the acceleration could be spurious. Elongated image?  \\
HIP 100934: data of low quality. \\
HIP 101726:  {\it first  resolution,} $P^*=70$\,yr, constant  RV. This
binary was  not resolved  by speckle interferometry,  possibly because
the companion is too faint in the visible. It is an X-ray source and a
``PMS star'' according to SIMBAD.  \\
HIP 102130:  unresolved. Spectroscopic binary according to both GCS and CfA. \\
HIP   103626:  {\it  first   resolution}  of   the  close   pair  with
acceleration. No RV data.  This is an X-ray source. \\
HIP 103983: {\it first resolution,}  $P^*=10$\,yr.  The pair can be followed
with  speckle because  of  small $\Delta  m$.  It is  on  the Keck  exo-planet search
program. \\
HIP 105872: {\it first resolution.} \\
HIP  105879: unresolved.  Variable  RV.  The  companion HJ~5267~AB  at
5$''$  is listed  in the  WDS  with one  measurement and  is not  seen
here. It should be considered spurious. \\
HIP  106560: unresolved.  Astrometric and  spectroscopic  binary. High
proper motion, low metallicity.  \\
HIP  107731:  the  companion  at 5\farcs6,  307$^\circ$  is  physical,
confirmed by 2MASS at 5\farcs3, 309$^\circ$ and hence CPM (the PM of A
is  0\farcs37  per  year).    The  companion's  photometry  in  2MASS,
$J-K=-0.15$ and $\Delta K=2.48$,  is uncertain and likely  distorted
by  the  proximity  of  A.  The  photometry  in  Table~1  is  reliable
(PSF-fitting) and  implies a red  companion color ($\Delta m$ of 3.6 and
4.2 in the read and  blue channels respectively).  This wide companion
cannot explain the  acceleration. \\
HIP 108589: unresolved.  With only a small $\Delta
\mu=7$\,mas~yr$^{-1}$ it can be single, despite two discordant RV
measures in the GCS. \\
HIP 109110: the spectroscopic binary with a preliminary 13-yr period
(CfA) and expected semi-major axis of 0\farcs19 is {\it tentatively
  resolved here for the first time} at 0\farcs07 with $\Delta K=1.7$.
Strangely, it was not resolved previously by speckle and by AO
\citep{MH09}. This is a young BY~Dra variable NT~Aqr and an X-ray
source.  \\
HIP 110340: unresolved.  With $\Delta \mu=7$\,mas~yr$^{-1}$ and
constant RV, it might be single. \\
HIP 113543: {\it first resolution} at 0\farcs21 is secure and hints at
a 30-yr period, but the preliminary SB orbit (CfA) has 9-yr period and
corresponds to 0\farcs09  semi-major axis.  Follow-up observations are
needed. \\
HIP 114880: tentatively resolved on 2011.85 at 147$^\circ$, 0\farcs10,
$\Delta  K=2.5$, but not  confirmed here.  As the  RV is  variable, we
suppose that the pair was  actually resolved last year, but has closed
in. \\
HIP 117247: {\it first resolution} at 0\farcs15, $P^* = 12$\,yr. \\
HIP  117258:  resolved  on  2011.85  and found  here  at  nearly  same
position, despite $P^*=20$\,yr. Seen in projection? \\
HIP 117475:  {\it first resolution} at 0\farcs20,  $\Delta K=3.2$. The
companion  is  seen  very   clearly,  of  nearly the same  magnitude  and
separation as the NICI ghost. No RV data. \\
HIP 118040: {\it first resolution}, $P^*=75$\,yr.}
\end{deluxetable*}

\end{document}